\documentclass[fleqn,10pt]{wlscirep}
\usepackage[utf8]{inputenc}
\usepackage[T1]{fontenc}
\usepackage{hyperref}
\hypersetup{
  colorlinks = true,
  citecolor = blue,
  urlcolor = blue
}

\usepackage{braket}
\usepackage{graphicx}
\usepackage[mathscr]{euscript}
\usepackage{amsmath}
\usepackage{amsfonts}
\usepackage{bm}
\usepackage{array}
\usepackage{float}

\newcolumntype{x}[1]{>{\hfil$\displaystyle} p{#1} <{$\hfil}}
\newcommand{{\bfL}}{\mbox{\boldmath$L$\unboldmath}}
\newcommand{{\bfl}}{\mbox{\boldmath$l$\unboldmath}}
\newcommand{{\bftheta}}{\mbox{\boldmath$\theta$\unboldmath}}
\newcommand{{\bfdelta}}{\mbox{\boldmath$\delta$\unboldmath}}
\newcommand{{\bfeta}}{\mbox{\boldmath$\eta$\unboldmath}}
\newcommand{{\bfbeta}}{\mbox{\boldmath$\beta$\unboldmath}}
\newcommand{{\bfphi}}{\mbox{\boldmath$\phi$\unboldmath}}
\newcommand{{\bfrho}}{\mbox{\boldmath$\rho$\unboldmath}}
\newcommand{{\bfcalB}}{\mbox{\boldmath$\cal B$\unboldmath}}
\newcommand{{\bfcalE}}{\mbox{\boldmath$\cal E$\unboldmath}}
\newcommand{{\bfcalJ}}{\mbox{\boldmath$\cal J$\unboldmath}}
\newcommand{{\bfcalO}}{\mbox{\boldmath$\cal O$\unboldmath}}
\newcommand{{\bfA}}{\mbox{\boldmath$\emph{A}$\unboldmath}}
\newcommand{{\bfB}}{\mbox{\boldmath$\emph{B}$\unboldmath}}

\def\v#1{{\bf#1}}

\DeclareMathAlphabet\mathbfcal{OMS}{cmsy}{b}{n}

\title{\LARGE{Can classical electrodynamics predict nonlocal effects?}}
\author[*]{\normalsize{Jos\'e A. Heras}}
\affil[*]{Instituto de Geof\'isica, Universidad Nacional Aut\'onoma de M\'exico, Ciudad de M\'exico 04510, M\'exico }

\author[**]{\normalsize{ Ricardo Heras}}
\affil[**]{Department of Physics and Astronomy, University College London, London WC1E 6BT, UK
 }

\begin{abstract}
Classical electrodynamics is a local theory describing local interactions between charges and electromagnetic fields and therefore one would not expect that this theory could predict nonlocal effects. But this perception implicitly assumes that the electromagnetic configurations lie in simply connected regions. In this paper we consider an electromagnetic configuration lying in a non-simply connected region, which consists of a charged particle encircling an infinitely-long solenoid enclosing a uniform magnetic flux, and show that the electromagnetic angular momentum of this configuration describes a nonlocal interaction between the encircling charge outside the solenoid and the magnetic flux confined inside the solenoid. We argue that the nonlocality of this interaction is of topological nature by showing that the electromagnetic angular momentum of the configuration is proportional to a winding number. The magnitude of this electromagnetic angular momentum may be interpreted as the classical counterpart of the Aharonov-Bohm phase.
\end{abstract}
\begin{document}

\flushbottom
\maketitle

\thispagestyle{empty}
\section{Introduction}
\noindent Three years after the publication of the seminal paper by Aharonov and Bohm\cite{1} on the prediction of the effect that now bears their name, the Aharonov-Bohm (AB) effect, Tassie and Peshkin\cite{2} claimed that ``... the quantum mechanical effects of the inaccessible field can be understood, both mathematically and physically, through angular momentum considerations ...'' They argued that in regions where electrons encircling inaccessible magnetic fields, classical physics predicts that, in addition to the mechanical angular momentum: $L_{\rm m}=(\v r\times m\v v)_z$, there is an electromagnetic angular momentum along the $z$-axis whose magnitude is given by
\begin{equation}
L_{\rm z}=\frac{e\Phi}{2\pi c},
\end{equation}
where $e$ is the electron charge and $\Phi$ is the inaccessible magnetic flux. The interesting point is that in these same free-field regions, quantum mechanics predicted the existence of the AB phase $\delta=e\Phi/(\hbar c)$. However, the comments of Tassie and Peshkin received little attention and the idea that (1) could shed light on the interpretation of the AB effect did not seem to have been explored further until in recent years in which Tiwari  \cite{3} and Wakamatsu et al.\cite{4}  derived (1) and drew attention to the role that (1) plays in the context of the AB effect.
In this paper we examine in detail the configuration formed by a charged particle encircling an infinitely-long solenoid enclosing a uniform magnetic flux and show that this particle  moves in a non-simply connected region where there is no magnetic field but  there is a non-zero vector potential. We then show that (1) is the electromagnetic angular momentum of this configuration. In contrast to Tiwari\cite{3} and Wakamatsu et al,\cite{4} who discuss (1) in connection to the AB effect, we only consider this connection when we suggest that (1) may be interpreted as the classical counterpart of the AB phase. Our main purpose here is to argue that (1)  describes a nonlocal interaction between the charge moving outside the solenoid and the magnetic flux inside this solenoid, which answers in the affirmative the question posed in the title of this paper. Throughout our study
we emphasize the topological and nonlocal features of (1).

\section{Vector potential of a closed flux line}
Consider  the charge-solenoid configuration, which consists of a particle having the charge $q$ and mass $m$ and continuously moving around an infinitely-long solenoid of radius $R$ which encloses a uniform magnetic flux $\Phi$. The $z$-axis is chosen as the axis of the solenoid (See Fig.~1). We use Gaussian units and cylindrical coordinates $(\rho,\theta,\phi)$ with their corresponding unit vectors $(\,\hat{\!\bfrho}, \hat{\!\bftheta},\hat{\!\bfphi})$. The electric current density of the infinitely-long solenoid is given by
\begin{equation}
\v J = \frac{c\Phi\delta(\rho-R)}{4\pi^2 R^2}\,\hat{\!\bfphi},
\end{equation}
where $\delta(\rho-R)$ is the Dirac delta function and $\Phi = \pi R^2 B$ is the magnetic flux through the solenoid with $B$ being the magnitude of the uniform magnetic field inside the solenoid. The electric current density $\v J$ is a steady current: $\nabla\cdot \v J=0$ and its magnetic field satisfies the magnetostatic equations
\begin{equation}
\nabla \cdot \v B =0,\quad \nabla \times\v B= \frac{\Phi\delta(\rho-R)}{\pi R^2}\,\hat{\!\bfphi},
\end{equation}
whose solution gives the explicit form of this magnetic field
\begin{equation}
\v B=\frac{\Phi \Theta(R-\rho)}{\pi R^2}\hat{\v z},
\end{equation}
which is  confined in the solenoid. Here $\Theta(\rho-R)$ is the Heaviside step function having the values $\Theta=1$ if $R>\rho$ and $\Theta=0$ if $R<\rho.$  To verify that (4) satisfies (3) we write $\v B= B_{z}(\rho)\hat{\v z}$, where $B_z(\rho)=\Phi \Theta(R-\rho)/(\pi R^2).$ Thus, $\nabla \cdot \v B= \partial B_z/\partial z=0$ and
$\nabla \times\v B =-(\partial B_z/\partial \rho)\,\hat{\!\bfphi}=\Phi\delta(\rho-R)\,\hat{\!\bfphi}/(\pi R^2),$ where we have used $\partial \Theta(R-\rho)/\partial \rho=-\delta(\rho-R)$. From (4) it follows that $\v B_{\rm out}=0$ and  $\v B_{\rm in}=\Phi\hat{\v z}/(\pi R^2),$
where $\v B_{\rm out}(\rho>R)$ and $\v B_{\rm in}(\rho<R)$ denote the values of the magnetic field  outside and inside the solenoid. The magnetic field vanishes outside the solenoid and has a constant value inside it. From the first equation in (3) we infer $\v B = \nabla \times \v A$ where $\v A$ is the corresponding vector potential. Using this relation in the second equation in (3), considering $\nabla^2\v F = \nabla(\nabla \cdot \v F)-\nabla \times (\nabla \times \v F)$ and adopting the Coulomb gauge $\nabla \cdot \v A=0$, we obtain the Poisson equation
\begin{equation}
\nabla^2 \v A = -\frac{\Phi\delta(\rho-R)}{\pi R^2}\,\hat{\!\bfphi}.
\end{equation}
In Appendix A we verify that the solution of this equation reads: \cite{5}
\begin{equation}
\v A= \frac{\Phi}{2 \pi}\bigg(\frac{\Theta(\rho-R)}{\rho}+\frac{\rho\,\Theta(R-\rho)}{R^2}\bigg)\,\hat{\!\bfphi}.
\end{equation}
This potential is not defined at $\rho=R$, i.e., at the surface of the solenoid. However, an appropriate regularisation yields $\v A(R)= \Phi\,\hat{\!\bfphi}/(2\pi R)$ which indicates that $\v A$ is continuous at $\rho=R.$
We can directly verify that (6) satisfies the Coulomb gauge: $\nabla \cdot \v A= (1/\rho)(\partial A_{\phi}/\partial \phi)=0.$ In Appendix B we verify that the curl of (6) yields the magnetic field given in (4).
\begin{figure}
	\centering
	\includegraphics[scale=0.3]{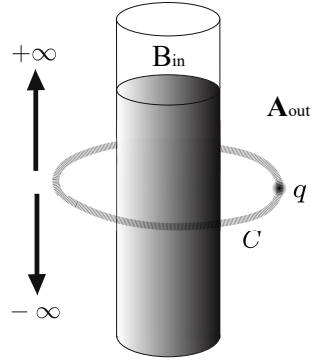}
	\caption{\small{Charge-solenoid configuration. An electric charge moving in the $x$-$y$ plane along the path $C$ which encircles an infinitely-long solenoid enclosing a uniform magnetic flux.}}
	\label{Fig1}
\end{figure}
Now, from (6) it follows that
\begin{equation}
\v A_{\rm out}= \frac{\Phi}{2 \pi\rho}\,\hat{\!\bfphi},\quad\v A_{\rm in}= \frac{\rho\,\Phi}{2 \pi R^2}\,\hat{\!\bfphi},
\end{equation}
where $\v A_{\rm out}(\rho>R)$ and $\v A_{\rm in}(\rho<R)$ denote the values of the vector potential outside and inside the solenoid. These potentials  are associated to the magnetic fields $\v B_{\rm out}=\nabla \times \v A_{\rm out}=0$ and $\v B_{\rm in}=\nabla \times \v A_{\rm in}=  \Phi\hat{\v z}/(\pi R^2).$ The first of these equations implies that $\v A_{\rm out}$  is a pure gauge potential: $\v A_{\rm out}= \nabla \chi,$
where the function $\chi=\chi(\phi)$ is defined as $\chi= \Phi \phi/(2\pi).$ This is a multi-valued function $\chi(\phi)\neq \chi(\phi+2 \pi)$ and satisfies $\nabla^2\chi=0.$ On the other hand, the circulation of the potential $\v A$ along an arbitrary closed path $C$ is invariant under the gauge transformation $\v A'=\v A+\nabla\Lambda$ where $\Lambda$ is a suitable gauge function. This gauge transformation must be a restricted gauge transformation because the potential $\v A$ is already in the Coulomb gauge. Therefore the function $\Lambda$ satisfies $\nabla^2\Lambda=0$. We also require that the function $\Lambda$ be single-valued. Under these considerations, the gauge invariance of the circulation of the vector potential $\v A$ follows
\begin{equation}
\oint_C \v A' \cdot d \v x= \oint_C \v A \cdot d \v x+ \oint_C \nabla\Lambda \cdot d \v x= \oint_C \v A \cdot d \v x.
\end{equation}
Here we have used the Stokes theorem $\oint_C \nabla\Lambda \cdot d \v x=\int_S \nabla \times \nabla\Lambda\cdot d \v S,$ where $S$ is the surface enclosed by the path $C$, and the identity $\nabla \times \nabla\Lambda = 0$ to conclude $\oint_C \nabla\Lambda \cdot d \v x=0.$ If we choose a path $C$ along a region inside the solenoid $\rho<R$, then $\v A'=\v A'_{\rm in}$ and from (8) it follows $\oint_C \v A'_{\rm in} \cdot d \v x= \oint_C \v A_{\rm in} \cdot d \v x.$ On the other hand, if the path $C$ lies in a region outside the solenoid $\rho>R$ then $\v A'=\v A'_{\rm out}$ and (8) yields $\oint_C \v A'_{\rm out} \cdot d \v x= \oint_C \v A_{\rm out} \cdot d \v x.$
We should stress that the result in (8) holds as long as the gauge function $\Lambda$ is singled-valued. If this is not true then we cannot assume the validity of (8) because the vanishing of the circulation $\oint_C \nabla\Lambda \cdot d \v x$ cannot generally be assumed (see Sec.~3).

The region outside the solenoid is shown to be an electromagnetic force-free region. To see this let us consider the Lorentz force $\mathbf{F} = q \dot{\v x}\times \v B(\v x)/c$, where $\dot{\v x}=d\v x/dt$ is the velocity of the charged particle and $\v x$ its position. Inserting the magnetic field (4) and $\dot{\v x} = \dot{\rho}\,\hat{\!\bfrho} + \rho \dot{\phi}\,\hat{\!\bfphi}+ \dot{z}\hat{\v z}$, we obtain the Lorentz force in the form
\begin{equation}
\v F = \frac{q\Phi \Theta(R-\rho)}{\pi R^2}(\rho \dot{\phi}\,\hat{\!\bfrho}-\dot{\rho}\,\hat{\!\bfphi}),
\end{equation}
which is local in the sense that the magnetic field must be evaluated at the point where the particle is located. It follows that if the particle is located inside the solenoid $\rho <R$ then $\Theta=1$ and we obtain $\v F = q\Phi(\rho \dot{\phi}\,\hat{\!\bfrho}-\dot{\rho}\,\hat{\!\bfphi})/(\pi R^2).$ However, in the charge-solenoid configuration the moving charged particle is assumed to be outside the solenoid $\rho>R$ so that $\Theta=0$ and therefore $\v F = 0.$ In short: the Lorentz force vanishes because the magnetic field is zero in the region where the charge is moving.

\section{Topology and nonlocality of the circulation of the vector potential outside the solenoid}
The Cauchy's integral formula \cite{6} for an analytic function $f(\textsc{z})$ reads,
\begin{eqnarray}
\frac{1}{2\pi i}\oint_C\frac{f(\textsc{z})\,d\textsc{z}}{\textsc{z}\!-\!\textsc{z}_0}=\left\{\begin{array}{@{}l@{\quad}l}
       n(C,\textsc{z}_0)f(\textsc{z}_0)\,& \mbox{if $C$ encloses encloses $\textsc{z}_0$} \\[\jot]
      \,0 & \mbox{otherwise}
    \end{array}\right.
\end{eqnarray}
where $n(C,\textsc{z}_0)$ represents the winding number of the curve $C$ around the singularity $\textsc{z}_0$. The winding number gives the number of times the curve $C$ encircles (counterclockwise) around a singularity. Consider now the particular case in which there is a singularity at $\rho =0,$ which in the complex plane corresponds to $\textsc{z}_0= \rho\,{\rm e}^{i\phi}=0$. We choose $f(\textsc{z})=K$, where $K$ is a constant and thus (10) yields
\begin{eqnarray}
\frac{K}{2 \pi i}\oint_C\frac{d\textsc{z}}{\textsc{z}}=\left\{\begin{array}{@{}l@{\quad}l}
       nK\,& \mbox{if $C$ encloses encloses $\textsc{z}_0=0$} \\[\jot]
      \,0 & \mbox{otherwise}
    \end{array}\right.
\end{eqnarray}
If we write  $\textsc{z}= \rho\, {\rm e}^{i \phi}$ and $d\textsc{z}= {\rm e}^{i\phi}(d\rho+i \rho d\phi)$ then it follows that  $d\textsc{z}/\textsc{z}=(1/\rho)d\rho+id\phi$. Therefore the left-hand side of (11) gives rise to two terms. The first term reads
$(K/2 \pi i)\oint_C d\rho/\rho$ but
$d \rho/\rho=(\hat{\rho}/\rho)\cdot d \v x= \nabla\ln{(\rho)}\cdot d \v x$, where $d \v x= d\rho\,\hat{\!\bfrho} + \rho d\phi\,\hat{\!\bfphi} + dz\hat{\v z}$ is the line element, implying $\oint_C\nabla\ln{(\rho)}\cdot d \v x=0$ because $\ln{(\rho)}$ is a single-valued function and consequently this first term vanishes. The second term is non-vanishing and reads $(K/2 \pi) \oint_C d\phi$. Since $d \phi=\,\hat{\!\bfphi}/\rho\cdot d \v x$ then (11) takes the particular form
\begin{eqnarray}
\oint_C \bigg(\frac{K\,\hat{\!\bfphi}}{2\pi\rho} \bigg)\cdot d \v x=\left\{\begin{array}{@{}l@{\quad}l}
       nK\,& \mbox{if $C$ encloses encloses $\rho=0$} \\[\jot]
      \,0 & \mbox{otherwise}
    \end{array}\right.
\end{eqnarray}
This is a purely formal result arising from the Cauchy's integral formula given in (10). Let us now apply (12) to
an idealised infinitely-long solenoid enclosing a uniform magnetic flux $\Phi$. We make the identification $K=\Phi$ in (12) and obtain the relation $K\,\hat{\!\bfphi}/(2\pi\rho)=\Phi\,\hat{\!\bfphi}/(2 \pi \rho)=\v A_{\rm out}(\v x)$. Since the solenoid encloses the singularity at $\rho=0$ then it follows
\begin{eqnarray}
\oint_C\!\v A_{\rm out}\!\cdot \!d \v x=\left\{\begin{array}{@{}l@{\quad}l}
       n\Phi & \mbox{if $C$ encloses the solenoid} \\[\jot]
      \,0 & \mbox{otherwise}
    \end{array}\right.
\end{eqnarray}
Here $n$ denotes the winding number of the curve $C$ around the solenoid. Equation (13) states that if the curve $C$ encircles $n$ times the solenoid then the circulation of the  potential $\v A_{\rm out}$  accumulates $n$ times the magnetic flux $\Phi$. The circulation of the vector potential in (13) is constant since $n\Phi$ is a constant quantity and therefore this circulation is insensitive to the form of the curve $C$ and also to the dynamics that we can associate to this curve. If for example we consider $C_1,C_2...C_k$ different curves which enclose counterclockwise $n$ times  the solenoid then (13) implies the equalities
\begin{eqnarray}
\oint_{C_1}\v A_{\rm out} \cdot d \v x = \oint_{C_2}\v A_{\rm out}\cdot \!d \v x=...=\oint_{C_k}\v A_{\rm out} \cdot d \v x.
\end{eqnarray}
The curves $C_1,C_2...C_k$ are homotopically equivalent (one curve can be continuously deformed into the other) and therefore we could not distinguish if the flux $\Phi$ in (13) is connected with the circulation of $\v A_{\rm out}$ along $C_1$ or along $C_2$ or along  $C_k$ (See Fig.~(2a)). This indistinguishability is a manifestation of the topology of the infinitely-long solenoid which lies in a non-simply connected region.
\begin{figure}
\centering
\includegraphics[scale=0.60]{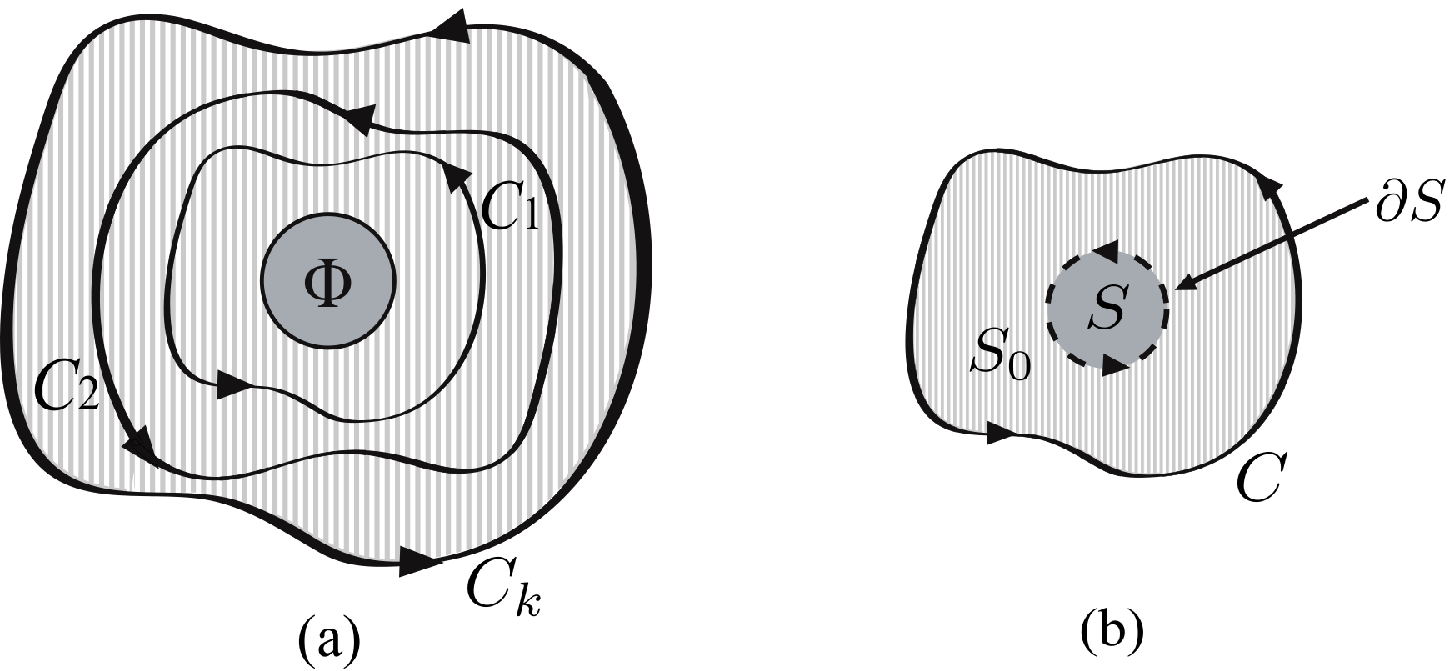}
\caption{(a) \small{The circulation of the vector potential $\v A_{\rm out}$ in a closed path around  the solenoid is insensitive to the form of the path, i.e., for different paths $C_1,C_2...C_k$ with the same winding number we have $\oint_{C_1}\!\v A_{\rm out}(\v x) \cdot d \v x\! =\! \oint_{C_2}\!\v A_{\rm out}(\v x) \cdot \!d \v x\!=...=\!\oint_{C_k}\!\v A_{\rm out}(\v x) \cdot d \v x.$ (b) The circulation $\oint_{C>\partial S}\v A_{\rm out}\cdot d \v x$ is taken along a path $C$ greater than the boundary $\partial S$ of the surface $S$ of the solenoid. Since $\oint_{C>\partial S}\v A_{\rm out}\cdot d \v x=\int_S \v B_{\rm in} \cdot d \v S$ holds then the circulation is spatially delocalised from the surface of the solenoid where the flux of the magnetic field is localised.}}
	\label{Fig2}
\end{figure}

Let us discuss the nonlocal feature of the circulation of the potential $\v A_{\rm out}$. At first sight, we could naively transform the closed line integral that defines this circulation into a surface integral via the Stokes theorem: $\oint_{C=\partial \mathscr{S}}\v A_{\rm out}\cdot d\v x= \int_{\mathscr{S}} \nabla\times \v A_{\rm out}\cdot d\v S,$ where the closed path $C$ now represents the boundary $\partial \mathscr{S}$ of some suitable surface $\mathscr{S}$. However, this application of the Stokes theorem leads to an inconsistent result because the potential $\v A_{\rm out}$ is the gradient of a function and therefore $\nabla\times \v A_{\rm out}=0$, which implies the vanishing of the circulation of $\v A_{\rm out}$, a result that contradicts (13) because it is assumed that the path $C$ encloses the solenoid. However, we can consistently use the Stokes theorem  but we need to be careful in applying it by taking into account that the path $C$ encloses $\rho=0$ and thus the considered region is non-simply connected and therefore $\oint_C\v A\cdot d\v x\not=0$.

Consider the vector potential $\v A$ in (6) which is defined in all space. Let us apply the Stokes theorem to this potential: $\oint_{C=\partial \mathscr{S}}\v A\cdot d\v x= \int_{\mathscr{S}} \nabla\times \v A\cdot d\v S.$ If the path $C$ encloses the solenoid then along this path the potential in (6) reads $\v A=\v A_{\rm out}$ and the Stokes theorem gives: $\oint_{C=\partial \mathscr{S}}\v A_{\rm out}\cdot d\v x= \int_{\mathscr{S}} \nabla\times \v A\cdot d\v S.$ In this case the surface $\mathscr{S}$ enclosed by the path $C$ can be written as ${\mathscr{S}}=S_{0} + S$, where $S_0$ is the surface extending outside the solenoid to the boundary $\partial \mathscr{S}$ and $S$ is the surface accounting for the cross section of the solenoid and having the boundary $\partial S$ (see Fig.~2(b)).  We observe that the boundary of the surface of the solenoid satisfies $C\! =\!\partial \mathscr{S}\!>\!\partial S$ because the path $C$ is outside the solenoid ($C>\partial S$ indicates that the length of the curve $C$ is greater than the length of the boundary $\partial S$ of the solenoid). Along the surface $S_0$ the potential in (6) reads $\v A=\v A_{\rm out}$ and along the solenoid surface $S$ it becomes $\v A=\v A_{\rm in}.$ Thus, the Stokes theorem gives: $\oint_{C=\partial\mathscr{S}}\v A_{\rm out}\cdot d\v x= \int_{S_0} \nabla\times \v A_{\rm out}\cdot d\v S + \int_{S} \nabla\times \v A_{\rm in}\cdot d\v S.$ The first term on the right vanishes because $\nabla\times \v A_{\rm out}=0$ and thus the Stokes theorem takes the form
\begin{equation}
\oint_{C>\partial S}\v A_{\rm out}\cdot d \v x =\int_{S}\nabla \times \v A_{\rm in}\cdot d\v S,
\end{equation}
which admits a nonlocal interpretation. While the left-hand side of (15) is defined outside the solenoid, its right-hand side is defined inside the solenoid. Put differently, the sides of (15) are defined in different spatial regions which implies a nonlocal connection between them. We can see that this nonlocality is a consequence of having applied the Stokes theorem in a non-simply connected region (we observe that $\oint_C\v A_{\rm out}\cdot d\v x =\oint_C\nabla \chi\cdot d\v x\not=0$) and ultimately is a consequence of the topology of the idealised solenoid. The genesis of this topology lies in the fact that in the assumption of an infinitely-long solenoid there is a non-removable singularity along the $z$-axis. Since $\nabla \times \v A_{\rm in}=\v B_{\rm in},$ with $\v B_{\rm in}$ denoting the constant magnetic field inside the solenoid, then (15) becomes
\begin{equation}
\oint_{C>\partial S}\v A_{\rm out}\cdot d \v x = \int_{S}\v B_{\rm in}\cdot d\v S,
\end{equation}
which expresses a nonlocal relation between the magnetic flux  confined in the solenoid and the circulation of the vector potential along a curve outside the solenoid. Equation (14) and (16) imply
\begin{equation}
\oint_{C_k>\partial S}\! \v A_{\rm out}\cdot d \v x=...=\oint_{C_2>\partial S} \!\v A_{\rm out}\cdot d \v x=\oint_{C_1>\partial S} \!\v A_{\rm out}\cdot d \v x=\int_{S} \v B_{\rm in} \cdot d \v S,
\end{equation}
Now, if we consider the equalities of the left-hand side of  the magnetic flux in Eq.~(17) then there is a manifest ambiguity because we cannot distinguish if this flux is connected with the circulation of $\v A_{\rm out}$ along $C_1>\partial S$ or along $C_2>\partial S$ or along  $C_k>\partial S$. This means that the circulations in (17) are spatially delocalised with respect to the magnetic flux, an expected result since they are not functions of point. Put in other words: the potential  $\v A_{\rm out}$ is ambiguous due to its gauge-dependence and its circulation $\oint_{C}\v A_{\rm out}\cdot d \v x$ is ambiguous due to its spatial delocalisation (indistinguishability of the curve $C$). We should also note the difference between applying the Stokes theorem in a simply connected region and in the non-simply connected region considered here. In the former application of the theorem there is a single curve $C=\partial S$ representing the boundary $\partial S$ of the surface $S$. In the latter application of the theorem there can be $k$ curves $C_k>\partial S$ all of them greater than the boundary $\partial S$ of the surface $S$.

We can now draw the lessons we have learned about the infinitely-long solenoid. The current  of this solenoid (2) yields the vector potential $\v A$ in (6) whose part outside the solenoid is $\v A_{\rm out}$ and the magnetic field $\v B$ in (4) whose part inside the solenoid is $\v B_{\rm in}$. This infinitely-long solenoid involves a line of singularity (along the $z$-axis) and then one can apply the Cauchy's integral formula (10) and the Stokes theorem (16) to this  non-simply connected region. Both mathematical tools lead to (17), which unambiguously shows the nonlocality of the circulation of $\v A_{\rm out}$ with respect to the flux of the magnetic field  $\v B_{\rm in}$. Physics deals with the infinitely-long solenoid and topology with the line of singularity and the homotopic curves around this line. One then can say that the idea of an infinitely-long solenoid is the genesis of the nonlocality  of the circulation of $\v A_{\rm out}$, or alternatively, that the idea of an infinite line of singularity is the genesis of this nonlocality. In more elegantly words: topology dictates nonlocality! However, an infinitely-long solenoid and a line of singularity are  abstract ideas and therefore the moral of this story is that it all started by making idealisations.

\section{Electromagnetic angular momentum}
Using the Poynting formula ${\bfL}_{\rm P}= [1/(4\pi c)]\int_{V}\mathbf{x} \times (\mathbf{E}\times \mathbf{B}) \,d^3x$ for an electromagnetic angular momentum in a volume $V$, we come directly to the conclusion that the electromagnetic angular momentum of the charge-solenoid configuration vanishes outside the solenoid because the magnetic field is zero in this region. However, Wakamatsu et al\cite{4} recently used the Maxwell formula:\cite{7} $\bfL_{\rm M}= (1/c)\int_{V}\varrho\,\v x \times \v A \,d^3x$ for an electromagnetic angular momentum and arrived at the conclusion that the electromagnetic angular momentum of the charge-solenoid configuration is given by $\bfL=q\Phi \hat{\textbf{z}}/(2\pi c)$ outside the solenoid. This same result was previously obtained by Tiwari\cite{3} using a Lagrangian treatment. Why do the Poynting  and Maxwell formulas predict different results for the same configuration?
To answer this question, let us formulate the following theorem:
\vskip 3pt
\noindent \emph{Decomposition Theorem.} Let $\mathbfcal{E}(\v x,t)$ be a time-dependent electric field and $\varrho(\v x,t)=\nabla\cdot \mathbfcal{E}/(4\pi)$ its associated charge density. Let $\v B(\v x)$ be a time-independent magnetic field and ${\v A}(\v x)$ its associated vector potential. The fields $\mathbfcal{E}$ and ${\v B}$ are independent fields, i.e., they are produced by different sources. The Poynting formula for the electromagnetic angular momentum in the volume $V$ originated by the interaction between the fields $\mathbfcal{E}$ and $\v B$ is given by
\begin{equation}
\textbf{\bfL}_{\rm P}=\frac{1}{4\pi c}\int_{V}\mathbf{x} \times (\mathbfcal{E}\times \mathbf{B})\,d^3x,
\end{equation}
and can be decomposed as
\begin{equation}
\textbf{\bfL}_{\rm P}=\textbf{\bfL}_{\rm M}+\textbf{\bfL}_{\rm R}+\textbf{\bfL}_{\rm G}+\textbf{\bfL}_{\rm S},
\end{equation}
where
\begin{align}
\textbf{\bfL}_{\rm M}=&\;\frac{1}{c}\int_{V}\varrho\,\mathbf{x} \times \mathbf{A}\,d^3x, \\
\textbf{\bfL}_{\rm R}=&\;\frac{1}{4\pi c}\int_{V}\mathbf{x} \times \big[(\nabla\times\mathbfcal{E})\times \v A\big]\,d^3x,\\
\textbf{\bfL}_{\rm G}=&\;\frac{1}{4\pi c}\int_{V}\mathbf{x} \times (\mathbfcal{E}\nabla\cdot \v A)\,d^3x,\\
\textbf{\bfL}_{\rm S}=&\;\frac{1}{4\pi c}\oint_{S} \mathbf{x} \times\big[\hat{\textbf{n}}(\mathbfcal{E}\cdot
\textbf{A})-\textbf{A}(\hat{\textbf{n}}\cdot\mathbfcal{E})-\mathbfcal{E}(\hat{\textbf{n}}\cdot \textbf{A})\big]\,d{S}.
\end{align}
Here $S$ is the surface of the volume $V$. The proof of this theorem is given in two parts. In  Appendix C we show that the validity of (18) follows from the Poynting vector associated to the interacting fields $\mathbfcal{E}$ and $\mathbf{B}$, and in Appendix D we show that (19) follows from a tensor identity.

The term $\textbf{\bfL}_{\rm R}$ in (21) contains the factor $\nabla\times\mathbfcal{E}$ which is connected with $(-1/c)\partial\mathbfcal{B}/\partial t$ due to the Faraday's induction law and therefore it deals with possible radiative effects. The term $\textbf{\bfL}_{\rm G}$ in (22) involving
$\nabla\cdot \v A$ deals with the adopted gauge for the potential $\v A$. The term $\textbf{\bfL}_{\rm S}$ in (23) is a surface term.
Notice that $\textbf{\bfL}_{\rm P}$ may be zero without $\textbf{\bfL}_{\rm M}$ necessarily be zero, which explains why the Poynting and Maxwell formulas can predict different results as it happens when considering the charge-solenoid configuration.

Let us now apply (19) to this charge-solenoid configuration where the volume $V$ covers all space except the volume of the solenoid. Here we have $\textbf{\bfL}_{\rm P}=0$ because $\textbf{B}=0$ outside the solenoid and therefore (19) yields
$0\!=\!\textbf{\bfL}_{\rm M}+\textbf{\bfL}_{\rm R}+\textbf{\bfL}_{\rm G}+\textbf{\bfL}_{\rm S}$. In Appendix E we show that
$\textbf{\bfL}_{\rm R}=0$. We have also  $\textbf{\bfL}_{\rm G}=0$ because $\nabla\cdot \v A_{\rm out}=0$. The surviving pieces satisfy  $0=\textbf{\bfL}_{\rm M}+\textbf{\bfL}_{\rm S}$ indicating that there exists $\textbf{\bfL}_{\rm M}$ in the volumetric space  where the charge is moving around the solenoid and there exists $\textbf{\bfL}_{\rm S}$ on the surface of this volumetric space which lies at infinity. It follows that $\textbf{\bfL}_{\rm S}=-\textbf{\bfL}_{\rm M}$ and this provides an interpretation for $\textbf{\bfL}_{\rm S}$.
We will see that $\textbf{\bfL}_{\rm M}$ is related to the flux of the magnetic field of the solenoid  and therefore $\textbf{\bfL}_{\rm S}$ may be interpreted as an electromagnetic angular momentum originated by the return flux of the magnetic field of an infinitely-long solenoid, an interpretation consistent with the fact that the signs of $\textbf{\bfL}_{\rm M}$ and $\textbf{\bfL}_{\rm S}$ are reversed. When considering the relation $\textbf{\bfL}_{\rm M}=-\textbf{\bfL}_{\rm S}$ we should always have in mind that $\textbf{\bfL}_{\rm M}$ and $\textbf{\bfL}_{\rm S}$ are defined in different spatial regions.

We also note that a different and less general two-dimensional decomposition of the Poynting formula $\bfL_{\rm P}$ was introduced by Wakamatsu et al \cite{4} (this decomposition contains a sign mistake), which involves a two-dimensional Coulomb field produced by the moving electron whose velocity is assumed to be ``much slower than the speed of light," a condition that, according to the authors, allows them to discard the magnetic field of the moving electron. They obtained the relation $\textbf{\bfL}_{\rm M}=-\textbf{\bfL}_{\rm S}$ and interpreted the piece $-\bfL_{\rm S}$ as an electromagnetic angular momentum originated by the return flux. However, their analysis is unsatisfactory because the electron-solenoid configuration is three-dimensional, and more importantly, because they completely ignored the radiative effects of the moving electron. Even for slowly moving electrons with $v\!<<\!c$ there are radiation fields as we will show it in Appendix E, where we will explicitly demonstrate the non-trivial result that the electromagnetic angular momentum of the charge-solenoid configuration is insensitive to radiative effects.

We now apply (20) to obtain the electromagnetic angular momentum of the charge-solenoid configuration covering all space except the surface $S$ which lies at infinity. We assume that the charge $q$ is localised in the $x$-$y$ plane so that its
position vector is $\v x_q(t)=\{\rho_q(t)\cos\phi_q(t),\rho_q(t)\sin\phi_q(t),0\}=\rho_q(t)\,\hat{\!\bfrho}$ and
$\varrho(\v x',t)=(q/\rho') \delta\{\rho' - \rho_q(t)\}\delta\{\phi' - \phi_q(t)\}\delta\{z'\}$ is its associated charge density. The vector potential is given by $\v A_{\rm out}(\v x')=\Phi\, \hat{\!\bfphi}/(2\pi\rho').$  A generic point reads $\v x'\! = \!\rho'\hat{\bfrho} \!+\!z'\hat{\v z}$. Using these ingredients and integrating the right-hand side of (20), we obtain
\begin{equation}
\frac{1}{c}\int_{V}\varrho(\v x',t)\, \v x' \times \v A_{\rm out}(\v x')\,d^3x'=\frac{n q \Phi}{2\pi c}\hat{\v z},
\end{equation}
where $n$ is the winding number identifying the number of times the electric charge travels its closed path around the solenoid. In getting (24) we have used  $\int^{\infty}_{-\infty}\delta(z')dz'=1,\int^{\infty}_{-\infty}z'\delta(z')dz'=0$,
\begin{equation}
\int^{\infty}_{R}\delta\{\rho'-\rho_q(t)\}\,d\rho'=\Theta\{\rho_q(t) - R\}=1,
\end{equation}
(because $\rho_q(t)>R$ outside the solenoid) and the relation
\begin{equation}
\oint_{C} \delta\{\phi' - \phi_q(t)\}\,d\phi'=n,
\end{equation}
which is demonstrated in Appendix F. Therefore (24) represents the accumulated electromagnetic angular momentum of the charge-solenoid configuration
\begin{equation}
\bfL=\frac{n q \Phi}{2\pi c}\hat{\v z}.
\end{equation}
Expressed differently, every time the charge goes around the solenoid, the charge-solenoid configuration acquires the electromagnetic angular momentum
$\bfL_{(n=1)}\!=\!q \Phi\hat{\v z}/(2\pi c)$ and therefore after $n$ times this configuration accumulates the electromagnetic angular momentum given by (27), which can compactly be written as $\bfL=n\bfL_{(n=1)}$ (this equation does not represent a quantisation rule because $\bfL_{(n=1)}$ does not generally describe a quantum of electromagnetic angular momentum).

Some additional comments on (27) are in order. The fact that this equation does not involve the coordinates of the encircling electric charge means that $\bfL$ is independent from the dynamics of this charge and therefore from its emitted radiation. This conclusion follows from the use of the relations (25) and (26) in the computation of (24). It is clear that the use of these integrals embodies a lost of information of the dynamical coordinates $\rho_q(t)$ and $\phi_q(t)$ of the encircling charge. Notice also that the corresponding electromagnetic angular momentum $\bfL_{\rm S}$ lying on the infinite surface and associated to the return flux, is given by
$\bfL_{\rm S}=-nq\Phi\hat{\mathbf{z}}/(2\pi c)$ and therefore we have the relation $\bfL_{\rm M}+\bfL_{\rm S}=0$ as predicted by the Poynting formula outside the solenoid. Here we justify the validity of this last relation by showing that $\textbf{\bfL}_{\rm P}=\textbf{\bfL}_{\rm M}+\textbf{\bfL}_{\rm R}+\textbf{\bfL}_{\rm G}+\textbf{\bfL}_{\rm S}$ with $\textbf{\bfL}_{\rm P}=0, \textbf{\bfL}_{\rm R}=0$ and $\textbf{\bfL}_{\rm G}=0$. On the other hand, when considering the Maxwell formula for the electromagnetic angular momentum applied to the charge-solenoid configuration
\begin{equation}
\bfL=\frac{1}{c}\int_{V}\varrho\, \v x \times \v A_{\rm out}\,d^3x,
\end{equation}
the question about its gauge invariance immediately arises. It is clear that this invariance requires $\int_{V} \varrho\,\v x \times\nabla \Lambda\,d^3x=0$ where $\Lambda$ is a suitable gauge function. This condition is valid for the charge-solenoid configuration. Let us express (28) in a form that makes evident its gauge invariance. We have seen that the relation $\oint_{C} \v A_{\rm out}\cdot d \v x\!=\!n\Phi$ given in (13) holds for the potential outside the infinite solenoid where $n$ is the winding number of the path $C$. This relation and (27) yield the novel form
\begin{equation}\,
\bfL=\frac{q\hat{\v z}}{2\pi c}\oint_{C} \v A_{\rm out}\cdot d \v x,
\end{equation}
which is gauge invariant on account of the gauge invariance of the circulation of the potential $\v A_{\rm out}.$ According to (29) the accumulated electromagnetic angular momentum depends on the circulation of the potential outside the solenoid.  Using (16) we can see that (29) can be expressed as
\begin{equation}
\bfL= \frac{q\hat{\v z}}{2\pi c}\int_{S}\v B_{\rm in}\cdot d\v S,
\end{equation}
which states that the accumulated electromagnetic angular momentum calculated in a volume $V$  outside the solenoid depends on the flux of the magnetic field inside the solenoid. At first glance, (29) and (30) suggest different interpretations of the physical origin of the electromagnetic angular momentum $\bfL$. Let us combine (29) and (30) to obtain the relation
\begin{equation}
\frac{q\hat{\v z}}{2\pi c}\oint_{C} \v A_{\rm out}\cdot d \v x=\bfL= \frac{q\hat{\v z}}{2\pi c}\int_{S}\v B_{\rm in}\cdot d\v S.
\end{equation}
The first interpretation, which we will call the $\bfA$-explanation, is supported by the first equality in (31), according to which the vector potential locally acts through its circulation on the charged particle originating the electromagnetic angular momentum $\bfL$. Argued differently, classical electrodynamics predicts the existence of an effect in the region outside the solenoid, the electromagnetic angular momentum $\bfL$, which is physically originated by the local action of an electromagnetic quantity on the charged particle. Now the only electromagnetic quantity function of point defined in that region is the potential  $\v A_{\rm out}$ and therefore this potential should produce $\bfL$. In short: $\v A_{\rm out}$ exists in each point of the trajectory of $q$ and therefore $\v A_{\rm out}$ locally acts on $q$ producing $\bfL$. The second interpretation, which we will call the $\bfB$-explanation, is supported by the second equality in (31), according to which the magnetic field nonlocally acts through its flux on the charged particle originating the electromagnetic angular momentum $\bfL$. In short: $\v B_{\rm in}$ exists inside the solenoid and not in each point of the trajectory of $q$ outside the solenoid and therefore $\v B_{\rm in}$ nonlocally acts on $q$ producing $\bfL$. We admit the $\bfB$-explanation and reject the $\bfA$-explanation for the following arguments:
\vskip 5pt
\noindent (a)	The potential $\v A_{\rm out}$ is gauge-dependent. 
\vskip 5pt
\noindent (b) The Stokes theorem applied in a non-simply connected region implies the relation (see (17))
\begin{equation}
\frac{q\hat{\v z}}{2\pi c}\oint_{C_k>\partial S}\! \v A_{\rm out}\cdot d \v x=...=\frac{q\hat{\v z}}{2\pi c}\oint_{C_2>\partial S} \!\v A_{\rm out}\cdot d \v x=\!\frac{q\hat{\v z}}{2\pi c}\oint_{C_1>\partial S} \!\v A_{\rm out}\cdot d \v x=\bfL=\frac{q\hat{\v z}}{2\pi c}\int_{S} \v B_{\rm in} \cdot d \v S,
\end{equation}
where the different charge paths $C_1>\partial S, C_2>\partial S, ..., C_k>\partial S$ possessing all of them the same winding number are homotopically equivalent. Now, if we consider the equalities of the left-hand side of  $\bfL$ in (32) then there is a manifest ambiguity because we cannot distinguish if $\bfL$ is connected with the circulation of $\v A_{\rm out}$ along $C_1>\partial S$ or along $C_2>\partial S$ or along  $C_k>\partial S$. We cannot know which of the circulations of the potential $\v A_{\rm out}$ displayed in (32) is causally connected with $\bfL$ because these circulations are spatially delocalised with respect to the solenoid (they are not functions of point). As pointed in Sec. 3, the potential  $\v A_{\rm out}$ is ambiguous due to its gauge-dependence and its circulation $\oint_{C}\v A_{\rm out}\cdot d \v x$ is ambiguous due to its spatial delocalisation (indistinguishability of the curve $C$). Accordingly, the $\bfA$-explanation does not hold. But if we consider the last equality in (32) then we conclude that $\bfL$ outside the solenoid is unambiguously connected with the flux of the magnetic field confined inside the solenoid. Since the charge and the magnetic flux lie in different spatial regions then they nonlocally interact to produce $\bfL$. Stated differently, (32) tell us that the magnetic flux does not locally affect
the trajectory of the charge.
\begin{figure}
	\centering
	\includegraphics[scale=0.75]{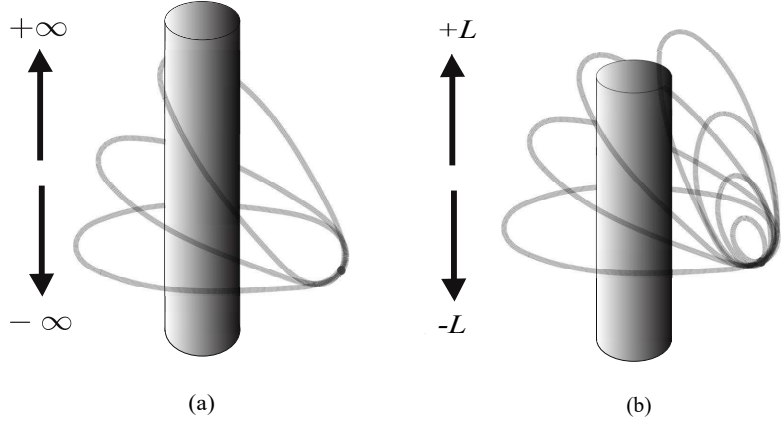}
	\caption{\small{(a) exhibits an idealised infinitely-long solenoid defined in a non-simply connected region and (b) exhibits a real finite solenoid defined in a simply connected region.}}
	\label{Fig3}
\end{figure}
We can now answer the question posed in the title of this paper:
Can classical electrodynamics predict non-local effects? Answer: yes, the electromagnetic relation given in (27) describes a nonlocal interaction between the magnetic flux confined inside the solenoid and the electric charge moving outside the solenoid. This electromagnetic configuration lies in a non-simply connected region. However, this kind of configurations is not usually considered in
the practice. The electromagnetic configurations we typically consider lie in simply-connected regions for which classical electrodynamics predicts local effects. Put differently, for the prediction of the electromagnetic angular momentum of the charge-solenoid configuration, it is crucial that the cylindrical solenoid be infinitely-long, which guarantees that the magnetic field is confined in this solenoid, that the associated vector potential outside the solenoid is a pure gauge potential and that this configuration lies in a non-simply connected region where curves encircling the infinitely-long solenoid cannot be shrunk (i.e., continuously deformed) into a point without overstepping this solenoid (see Fig.~3(a) where we have drawn curves in space not in the plane with the purpose of making the argument clearer). If the cylindrical solenoid were long but finite then
the magnetic field is no longer confined in the solenoid, the associated vector potential outside the solenoid is not a pure gauge potential, and this configuration lies in a simply connected region where curves encircling this solenoid can be shrunk into a point bypassing this solenoid (see Fig.~3(b)).

\section{The classical counterpart of the AB phase}
We have argued that classical electrodynamics and the topology of the charge-solenoid configuration conspire to produce the accumulated electromagnetic angular momentum of magnitude
\begin{equation}
L=\frac{q}{2\pi c}\oint_{C>\partial S}\v A_{\rm out}\cdot d \v x = \frac{q}{2\pi c}\int_{S}\v B_{\rm in}\cdot d\v S=\frac{nq\Phi}{2\pi c},
\end{equation}
In a similar fashion, quantum mechanics and the topology of the charge-solenoid configuration also conspire to produce the accumulated quantum AB phase
\begin{equation}
\delta=\frac{q}{\hbar c}\oint_{C>\partial S}\v A_{\rm out}\cdot d \v x = \frac{q}{\hbar c}\int_{S}\v B_{\rm in}\cdot d\v S=\frac{nq\Phi}{\hbar c}.
\end{equation}
Both quantities $L$ and $\delta$  satisfy the little-known linear relation: \cite{4}
\begin{equation}
\delta =\frac{2\pi}{\hbar}L,
\end{equation}
which expresses a striking connection between classical electrodynamics represented by the electromagnetic angular momentum $L$, and quantum mechanics represented by the AB phase $\delta$. As may be seen, both $L$ and $\delta$ are given in terms of the circulation of $\textbf{A}_{\rm out}$, which in turn satisfies the topological feature expressed in (13) and the nonlocal feature expressed in (16). The conclusion naturally arises: the topological and nonlocal features of the circulation of the vector potential outside the solenoid are  naturally translated to both the classical quantity $L$ and the quantum quantity $\delta$ and this result leads us to suggest that $L$ may be naturally considered the classical counterpart of $\delta$. Regarding this suggestion, it is pertinent to note that some authors have claimed that the AB phase has no classical counterpart  (see, for example, Refs.~8-10). However, a number of authors have proposed classical analogues of the AB  effect (see, for example, Refs.~11-20).

\section{Conclusion}
In this paper we have shown that classical electrodynamics can predict nonlocal effects even though this theory is the prototype of a local theory.
This nonlocality was revealed when we considered the configuration formed by a charged particle continuously encircling an infinitely-long solenoid enclosing a uniform magnetic flux. The charged particle moves in a non-simply connected region where there is no magnetic field and therefore there is no Lorentz force acting on the charge but there is a non-zero vector potential. We then derived the electromagnetic angular momentum of this configuration $\bfL=[q\hat{\v z}/(2\pi c)]\oint_{C>\partial S}\textbf{A}_{\rm out}\cdot d\v x=[q\hat{\v z}/(2\pi c)]\int_{S}\v B_{\rm in}\cdot d\v S=nq\Phi/(2\pi c),$ which is topological because it depends on the winding number $n$  representing the number of times the charge carries out around the solenoid and is nonlocal because the magnetic field of the solenoid acts on the charged particle in regions for which this field does not exist.
We excluded the idea that this electromagnetic angular momentum was physically produced by the vector potential because this potential is ambiguous due to its gauge-dependence and because the circulation of this potential is also ambiguous due to its spatial delocalisation. We have noted that the genesis of the topology and nonlocality of the obtained electromagnetic angular momentum is the idealised infinitely-long solenoid involving
 an infinite line of singularity, which allowed us to apply the Cauchy's integral formula and the Stokes theorem in a non-simply connected region.
Finally, we have argued that the magnitude of the derived electromagnetic angular momentum may be interpreted as the classical counterpart of the AB phase.

\renewcommand\theequation{A\arabic{equation}}
\setcounter {equation}{0}
\section*{Appendix A Proof that (6) satisfies (5)}
The Laplacian of an arbitrary vector of the form $\v F=F_{\phi}(\rho)\,\hat{\!\bfphi}$  is given by $\nabla^2 \v F = (\nabla^2 A_{\phi}  - A_{\phi}/\rho^2)\,\hat{\!\bfphi}$ in cylindrical coordinates. Equation (6) is of the form $\v A=A_{\phi}(\rho)\,\hat{\!\bfphi}$ with $A_{\phi}(\rho)$ given by
\begin{equation}
A_{\phi}(\rho)=\frac{\Phi}{2 \pi}\bigg(\frac{\Theta(\rho-R)}{\rho}+\frac{\rho\,\Theta(R-\rho)}{R^2}\bigg).
\end{equation}
The Laplacian of $A_{\phi}(\rho)$ reads
\begin{equation}
\nabla^2  A_{\phi} =  \frac{1}{\rho}\frac{\partial}{\partial \rho}\bigg( \rho\frac{\partial A_{\phi}}{\partial \rho}\bigg).
\end{equation}
Therefore
\begin{align}
\nonumber \rho\frac{\partial A_{\phi}}{\partial \rho}=&\, \frac{\rho\,\Phi}{2\pi}\bigg[\frac{\partial }{\partial \rho}\bigg(\frac{\Theta(\rho-R)}{\rho} \bigg) + \frac{1}{R^2}
\frac{\partial }{\partial \rho} \bigg( \rho \,\Theta(R-\rho)\bigg)\bigg]\\
\nonumber= &\,\frac{\rho\,\Phi}{2\pi}\bigg[ \bigg( \frac{\delta(\rho-R)}{\rho}-\frac{\Theta(\rho-R)}{\rho^2}\bigg)+\bigg( \frac{\Theta(R-\rho)}{R^2}-\frac{\rho\,\delta(R-\rho)}{R^2}\bigg)\bigg]\\
\nonumber=& \, \frac{\Phi}{2\pi}\bigg[ \bigg( \frac{\rho\, \Theta(R-\rho)}{R^2}-\frac{\Theta(\rho-R)}{\rho}\bigg)+  \frac{1}{R^2} \bigg((R^2-\rho^2)\delta(R-\rho)\bigg)\bigg]\\
=&\, \frac{\Phi}{2\pi}\bigg[ \frac{\rho\, \Theta(R-\rho)}{R^2}-\frac{\Theta(\rho-R)}{\rho}\bigg],
\end{align}
where it is evident that $(R^2-\rho^2)\delta(R-\rho)=0$ for $R>\rho$ and $R<\rho$. It follows
\begin{align}
\nonumber \nabla^2  A_{\phi} =& \, \frac{\Phi}{2\pi \rho}\bigg[\frac{\partial}{\partial \rho}\bigg( \frac{\rho\, \Theta(R-\rho)}{R^2}-\frac{\Theta(\rho-R)}{\rho}\bigg)\bigg]\\
\nonumber=&\, \frac{\Phi}{2\pi \rho}\bigg[  \bigg( \frac{\Theta(\rho-R)}{\rho^2}+ \frac{\Theta(R-\rho)}{R^2} - \frac{\rho\,\delta(R-\rho)}{R^2}-\frac{\delta(\rho-R)}{\rho} \bigg)\bigg]\\
\nonumber=&\, \frac{1}{\rho^2}\bigg[\frac{\Phi}{2\pi}\bigg( \frac{\Theta(\rho\!-\!R)}{\rho}\!+\! \frac{\rho\Theta(R\!-\!\rho)}{R^2} \bigg)\bigg]\!-\!\frac{\Phi}{2\pi}\bigg[\frac{\delta(R\!-\!\rho)}{R^2}\!+\!\frac{\delta(R\!-\!\rho)}{\rho^2} \bigg]\\
=&\,\frac{A_{\phi}}{\rho^2}-\frac{\Phi}{2\pi}\bigg[\frac{\delta(R\!-\!\rho)}{R^2}\!+\!\frac{\delta(R\!-\!\rho)}{\rho^2} \bigg].
\end{align}
Now, we can write
\begin{equation}
\frac{\delta(R-\rho)}{R^2}+\frac{\delta(R-\rho)}{\rho^2}=\, \frac{2\delta(\rho - R)}{R^2} + \frac{1}{R^2\rho^2}\bigg( (R^2-\rho^2) \delta(R-\rho) \bigg)=\,  \frac{2\delta(\rho - R)}{R^2},
\end{equation}
where we have considered the result $(R^2-\rho^2) \delta(R-\rho)=0$. Therefore
\begin{equation}
\nabla^2  A_{\phi} =\frac{A_{\phi}}{\rho^2} - \frac{\Phi\delta(\rho - R)}{\pi R^2},
\end{equation}
which implies the Poisson equation given in (5),
\begin{equation}
\nabla^2 \v A = \bigg( \nabla^2 A_{\phi}  - \frac{A_{\phi} }{\rho^2} \bigg)\,\hat{\!\bfphi}= -\frac{\Phi\delta(\rho - R)}{\pi R^2}\,\hat{\!\bfphi},
\end{equation}

\renewcommand\theequation{B\arabic{equation}}
\setcounter {equation}{0}
\section*{Appendix B Proof that the curl of (6) gives (4) }
 The curl of an arbitrary vector of the form $\v F=F_{\phi}(\rho)\,\hat{\!\bfphi}$ is given by $\nabla \times \v F= (1/\rho)( \partial(\rho F_{\phi})/\partial \rho)\hat{\v z}$ in cylindrical coordinates. It follows that
\begin{align}
\nabla\times\bigg(\frac{\Phi}{2 \pi\rho}\Theta(\rho-R)\,\hat{\!\bfphi}\bigg)=&\;\frac{\Phi}{2 \pi\rho}\delta(\rho-R)\hat{\v z}\\
\nabla\times\bigg(\frac{\Phi\rho}{2 \pi R^2}\Theta(R-\rho)\,\hat{\!\bfphi}\bigg)=&\;\frac{\Phi}{\pi R^2}\Theta(R-\rho)\hat{\v z}-\frac{\Phi\rho}{2\pi R^2}\delta(R-\rho)\hat{\v z}.
\end{align}
Therefore the curl of (6) is
\begin{align}
\v B=\nabla\times\bigg(\frac{\Phi}{2 \pi\rho}\Theta(\rho-R)\,\hat{\!\bfphi}+\frac{\Phi\rho}{2 \pi R^2}\Theta(R-\rho)\,\hat{\!\bfphi}\bigg)
=\frac{\Phi}{\pi R^2}\Theta(R-\rho)\hat{\v z}+\frac{\Phi}{2\pi}\bigg(\frac{R^2-\rho^2}{\rho R}\bigg)\delta(R-\rho)\hat{\v z}.
\end{align}
A non-vanishing value of $\v B$ occurs when $R> \rho$ is assumed in the first term but then the last term identically vanishes. Therefore the validity of (4) is demonstrated.

\renewcommand\theequation{C\arabic{equation}}
\setcounter {equation}{0}
\section*{Appendix C Poynting theorem of combined electromagnetic equations and proof of (18)}
Consider two independent systems of electromagnetic equations. The first system describes the time-dependent electric and magnetic fields ${\mathbfcal{E}}(\v x,t)$ and ${\mathbfcal{B}}(\v x,t)$ produced by an arbitrarily moving charge $q$ having the velocity $\dot{\v x}_q(t)= d \v x_q(t)/dt$ and position $\v x_q(t)$. The corresponding charge and current densities of the moving charge are given by $\varrho(\v x,t)= q \delta\{\v x- \v x_q(t)\}$ and $\mathbfcal{J}(\v x,t)=q\dot{\v x}_q(t) \delta\{\v x- \v x_q(t)\}$ which satisfy the continuity equation $\nabla \cdot \mathbfcal{J}+\partial\varrho/\partial t=0$ and the Maxwell equations
\begin{eqnarray}
\nabla\cdot{\mathbfcal{E}}=\,4\pi\varrho,\;\; \nabla\times{\mathbfcal{E}}+\frac{1}{c} \frac{\partial{\mathbfcal{B}}}{\partial t} =0,\;\;
\nabla\cdot{\mathbfcal{B}}=\,0, \;\;\nabla\times{\mathbfcal{B}}-\frac{1}{c} \frac{\partial{\mathbfcal{E}}}{\partial t} =\frac{4\pi}{c}\mathbfcal{J}.
\end{eqnarray}
The charged particle interacts with external electric and magnetic fields $\v E(\v x,t)$ and $\v B(\v x,t)$ produced in turn by their own set of charge and current densities $\rho(\v x,t)$ and $\v J(\v x,t)$ satisfying the continuity equation $\nabla \cdot \v J+\partial\rho/\partial t=0$ as well as the Maxwell equations
\begin{eqnarray}
 \nabla\cdot\v E=\,4\pi\rho,\;\; \nabla\times\v E+\frac{1}{c} \frac{\partial\v B}{\partial t} =0,\;\;
\nabla\cdot\v B=\,0,\;\; \nabla\times \v B-\frac{1}{c} \frac{\partial \v E}{\partial t} =\frac{4\pi}{c}\v J,
\end{eqnarray}
which constitute the second system of electromagnetic equations we are considering. We shall now obtain the corresponding Poynting theorem of the combined system formed by (C1)  and (C2). We start by considering the work done on the charged particle which, according to the Lorentz force, is
\begin{equation}
\v F \cdot d \v x_q= q\bigg(\v E(\v x_q, t) \!+\! \frac{\dot{\v x}_q(t)}{c} \times \v B(\v x_q, t) \bigg)\!\cdot\! \dot{\v x}_q(t)dt=q\v E(\v x_q, t)\!\cdot\! \dot{\v x}_q(t)dt.
\end{equation}
For the charged particle we may write $q=\rho(\v x_q,t)d^3x$ and $\dot{\v x}_q(t)\rho(\v x_q,t)=\mathbfcal{J}(\v x_q,t)$ so that the rate at which work is done on the charge within a volume $V$ reads
\begin{equation}
\int_V \v E\cdot \mathbfcal{J}\,d^3x.
\end{equation}
Our job now consist in casting the integrand of (C4) in the form of a local conservation law, i.e., to find something of the form $\nabla \cdot (...) + \partial(...)/\partial t,$ which is conserved in absence of sources. From the Ampere-Maxwell law displayed in (C1) we obtain $\mathbfcal{J} = (c/4\pi)\nabla \times \mathbfcal{B} -(1/4\pi)(\partial \mathbfcal{E}/\partial t)$ so that
\begin{equation}
\v E\cdot \mathbfcal{J}= \frac{c}{4\pi}\v E\cdot(\nabla \times \mathbfcal{B})- \frac{1}{4\pi} \v E\cdot \frac{\partial \mathbfcal{E}}{\partial t}.
\end{equation}
Using the identity $\nabla \cdot (\v E \times \mathbfcal{B})= \mathbfcal{B}\cdot(\nabla \times \v E)-\v E\cdot (\nabla \times \mathbfcal{B})$ and the Faraday law $\nabla \times \mathbfcal{E} = -(1/c)\partial \mathbfcal{B}/\partial t$ in (C1) we obtain
\begin{equation}
\v E \cdot \mathbfcal{J}=- \frac{1}{4\pi}\bigg(\mathbfcal{B} \cdot \frac{\partial \v B}{\partial t} + \v E\cdot \frac{\partial \mathbfcal{E}}{\partial t}\bigg)- \frac{c}{4\pi}\nabla \cdot(\v E\times \mathbfcal{B}).
\end{equation}
Since $\partial (\mathbfcal{B} \cdot \v B)/\partial t = \mathbfcal{B} \cdot (\partial \v B/\partial t)+ \v B\cdot (\partial \mathbfcal{B}/\partial t)$ and $\partial (\mathbfcal{E} \cdot \v E)/\partial t = \mathbfcal{E} \cdot (\partial \v E/\partial t)+ \v E\cdot (\partial \mathbfcal{E}/\partial t)$ it follows
\begin{equation}
 \v E\cdot \mathbfcal{J}=- \frac{1}{4\pi} \frac{\partial}{\partial t}\bigg( \mathbfcal{E}\cdot \v E+ \mathbfcal{B}\cdot \v B \bigg) + \frac{1}{4\pi}\bigg(\mathbfcal{E} \cdot \frac{\partial \v E}{\partial t} + \v B\cdot \frac{\partial \mathbfcal{B}}{\partial t} \bigg)- \frac{c}{4\pi}\nabla \cdot(\v E\times \mathbfcal{B}).
\end{equation}
Let us analyse the second term of (C7). Using the Faraday law $\partial \mathbfcal{B}/\partial t= -c\nabla \times \mathbfcal{E}$ in (C1) it follows the relation $\v B\cdot(\partial \mathbfcal{B}/\partial t)=-c\mathbfcal{E} \cdot(\nabla \times \v B)-c\nabla \cdot(\mathbfcal{E}\times \v B)$. Now, the Ampere-Maxwell law in (C2) implies
$\partial\v E/\partial t=-4\pi \v J + c \nabla \times \v B$ so that $\mathbfcal{E} \cdot (\partial \v E/\partial t)= - 4\pi \mathbfcal{E} \cdot \v J + c\mathbfcal{E} \cdot (\nabla \times \v B).$ Therefore
\begin{equation}
\frac{1}{4\pi}\bigg(\mathbfcal{E} \cdot \frac{\partial \v E}{\partial t} + \v B\cdot \frac{\partial \mathbfcal{B}}{\partial t} \bigg)=-\mathbfcal{E}\cdot \v J- \frac{c}{4\pi}\nabla \cdot(\mathbfcal{E}\times \v B),
\end{equation}
which is used in (C7) to obtain the relation
\begin{equation}
\v E\cdot \mathbfcal{J} = -\mathbfcal{E}\cdot \v J-\frac{1}{4\pi} \frac{\partial}{\partial t}\bigg( \mathbfcal{E}\cdot \v E + \mathbfcal{B}\cdot \v B \bigg)- \frac{c}{4\pi}\nabla\cdot (\v E\times \mathbfcal{B} + \mathbfcal{E}\times \v B).
\end{equation}
Let us now define the interaction energy density as
\begin{equation}
U= \frac{1}{4\pi}\bigg( \mathbfcal{E}\cdot \v E + \mathbfcal{B}\cdot \v B\bigg),
\end{equation}
and the interaction Poynting vector as
\begin{equation}
\v S= \frac{c}{4\pi}(\v E\times \mathbfcal{B} + \mathbfcal{E}\times \v B).
\end{equation}
Inserting (C9) into (C4) and using (C10) and (C11) we obtain
\begin{equation}
-\int_V (\v E\cdot \mathbfcal{J} + \mathbfcal{E}\cdot \v J)\,d^3x=\int_V \bigg(\frac{\partial U}{\partial t}+ \nabla \cdot \v S  \bigg) d^3x.
\end{equation}
Since the volume $V$ is arbitrary then the integrand of (C12) is valid for any point so that the Poynting theorem for the energy conservation due to the interaction of a charged particle with external electromagnetic fields reads
\begin{equation}
\nabla \cdot \v S + \frac{\partial U}{\partial t}=-\v E\cdot \mathbfcal{J} - \mathbfcal{E}\cdot \v J.
\end{equation}
Let us now to apply this interaction Poynting theorem to the charge-solenoid configuration where the charged particle satisfies (C1), there is the magnetostatic field ${\v B}(\v x)$ produced by the current density ${\v J}(\v x)$ playing the role of external magnetic field and there is no electric field. The magnetic field obeys the equations $\nabla\cdot{\v B}=0, \nabla\times{\v B}=4\pi \v J/c.$ The first equation is satisfied by $\v B=\nabla\times\v A,$ where $\v A(\v x)$ is the associated vector potential. For this particular case (C13) describes the interaction of the fields  ${\mathbfcal{E}}$ and  ${\v B}$ and takes the form $\nabla\cdot\v S+\partial U/\partial t=-{\mathbfcal{E}}\cdot\v J,$ where the Poynting vector is $\v S=c\,{\mathbfcal{E}}\times\v B/(4\pi)$
and the energy density is $U={\mathbfcal{B}}\cdot\v B/(4\pi)$. The corresponding  electromagnetic momentum density reads $ \v g={\mathbfcal{E}}\times\v B/(4\pi c)$
and its associated electromagnetic angular momentum density is given by $\mathbfcal{L}=\v x\times({\mathbfcal{E}}\times\v B)/(4\pi c),$ whose volume integration yields the electromagnetic angular momentum given in (18).
\renewcommand\theequation{D\arabic{equation}}
\setcounter{equation}{0}
\section*{Appendix D Proof of (19)}
A direct vector calculation leads to the identity
\begin{align}
\nonumber \partial^k\big[\epsilon^{sqi}x_q(\delta_{ik}{\cal E}_m A^m-{\cal E}_k A_i-{\cal E}_i A_k)\big]=&\;-4\pi\varrho\,\varepsilon^{sqi}x_q A_i+\varepsilon^{sqi}x_q {\cal E}^k(\partial_k A_i-\partial_i A_k)+\varepsilon^{sqi}x_q A^k(\partial_i{\cal E}_k-\partial_k{\cal E}_i)-\varepsilon^{sqi}x_q{\cal E}_i\partial^kA_k\nonumber \\
=&\;-4\pi\big[\varrho\,\v x\times \v A \big]^s +\big[\v x\times(\mathbfcal{E}\times\v B) \big]^s-\big[\v x\times\{(\nabla\times\mathbfcal{E})\times\v A\} \big]^s-\big[\v x\times\mathbfcal{E}(\nabla\cdot\v A)\big]^s.
\end{align}
Volume integration of (D1) implies the relation
\begin{align}
\frac{1}{4\pi c}\int_{V}\,[\mathbf{x} \times (\mathbfcal{E}\times \mathbf{B})]^s\,d^3x=&\;\frac{1}{c}\int_{V}[\varrho\,\mathbf{x} \times \mathbf{A}]^s\,d^3x+\frac{1}{4\pi c}\int_{V}[\mathbf{x} \times \big\{(\nabla\times\mathbfcal{E})\times \v A\big\}]^sd^3x+\frac{1}{4\pi c}\int_{V}[\mathbf{x} \times (\mathbfcal{E}\nabla\cdot \v A)]^s\,d^3x\nonumber\\&\;+\frac{1}{4\pi c}\oint_{S} \big\{\mathbf{x} \times\big[\hat{\textbf{n}}(\mathbfcal{E}\!\cdot\!
\textbf{A})-\textbf{A}(\hat{\textbf{n}}\!\cdot\!\mathbfcal{E})-\mathbfcal{E}(\hat{\textbf{n}}\cdot \textbf{A})\big]\big\}^sd{S}.
\end{align}
Notice that the volume integral of the left-hand side of (D1) has been transformed into a surface integral by making the replacements $\partial^k\to (\hat{\v n})^k$ and $\int_{V}d^3x\to\oint_{S}dS$. Clearly, (D2) implies (19).

\renewcommand\theequation{E\arabic{equation}}
\setcounter{equation}{0}
\section*{Appendix E Vanishing of $\bfL_{\rm R}$ for the charge-solenoid configuration }
 Using the Faraday's induction law $\nabla\times \mathbfcal{E}=-(1/c)\partial \mathbfcal{B}/\partial t$, (19) takes the form
\begin{equation}
\textbf{\bfL}_{\rm R}=-\frac{1}{4\pi c^2}\frac{\partial}{\partial t}\int_{V}\mathbf{x} \times \!\big(\mathbfcal{B}\times \v A\big)\,d^3x.
\end{equation}
Let us now apply this formula to the particular case of the charge-solenoid configuration by writing down $\v x\! = \!\rho\hat{\bfrho} \!+\!z\hat{\v z}$ and $\v A=\v A_{\rm out}(\v x)=\Phi\, \hat{\!\bfphi}/(2\pi\rho)$. It follows that
\begin{align}
-\frac{1}{4\pi c^2}\frac{\partial}{\partial t}\int_{V}\mathbf{x} \times \!\big(\mathbfcal{B}\times \v A)\,d^3x
=&\;-\frac{\Phi}{8\pi^2 c^2}\frac{\partial}{\partial
 t}\int_{V}\rho\,\hat{\!\bfrho}\times\!\big(\mathbfcal{B}\times\,\hat{\!\bfphi}\big)d\rho d\phi dz-\frac{\Phi}{8\pi^2 c^2}\frac{\partial}{\partial t}\int_{V}z\hat{\v z}\times\!\big(\mathbfcal{B}\times\,\hat{\!\bfphi}\big)d\rho d\phi dz.
\end{align}
The magnetic field $\mathbfcal{B}={\cal B}_\rho\,\hat{\!\bfrho}+{\cal B}_\phi\,\hat{\!\bfphi}+{\cal B}_z\hat{\v z}$  is a Lienard-Wiechert field produced by the encircling charge, which can be decomposed as $\mathbfcal{B}=\mathbfcal{B}^v+\mathbfcal{B}^a$, where $\mathbfcal{B}^v$ is a velocity field varying like $1/R^2$ and
$\mathbfcal{B}^a$ is an acceleration field varying like $1/R$. Using this decomposition in (E2) and performing the specified operations, we obtain
\begin{align}
-\frac{1}{4\pi c^2}\frac{\partial}{\partial t}\int_{V}\mathbf{x} \times\! \big(\mathbfcal{B}\times \v A)\,d^3x=&\;-\frac{\Phi\,\,\hat{\!\bfphi}}{8\pi^2 c^2}\frac{\partial}{\partial t}\Bigg[\int_{V}\rho{\cal B}^v_\rho\, d\rho d\phi dz
+ \int_{V}\rho{\cal B}^a_\rho\, d\rho d\phi dz\Bigg]\nonumber\\
&\;-\frac{\Phi\,\,\hat{\!\bfphi}}{8\pi^2 c^2}\frac{\partial}{\partial t}\Bigg[\int_{V}z{\cal B}^v_z\, d\rho d\phi dz
+\int_{V}z{\cal B}^a_z\, d\rho d\phi dz \Bigg].
\end{align}
The components ${\cal B}^v_\rho, {\cal B}^a_\rho, {\cal B}^v_z$ and ${\cal B}^a_z$ should be now specified. To simplify the problem, let us assume that the charged particle is moving with a non-relativistic velocity and consider that the charge moves in a circle of radius $a$ outside the solenoid at constant angular velocity $w$. The required components of the magnetic field $\mathbfcal{B}$ read
\begin{align}
{\cal B}^v_\rho=&\;0,\\
{\cal B}^a_\rho=&\;\frac{qw^2az\sin{(\phi-wt_{\rm r}})}
{c^2[\rho^2+z^2+a^2-2a\rho\cos{(\phi-wt_{\rm r}})]},\\
{\cal B}^v_z=&\;-\frac{qwa[\rho\cos{(\phi-wt_{\rm r}})-a]}
{\rho^2+z^2+a^2-2a\rho\cos{(\phi-wt_{\rm r}})},\\
{\cal B}^a_z=&\;\frac{qw^3a^2[\rho\cos{(\phi-wt_{\rm r}})-a]^2}
{c^2[\rho^2+z^2+a^2-2a\rho\cos{(\phi-wt_{\rm r}})]^{3/2}}-\frac{qw^2a\rho\sin{(\phi-wt_{\rm r}})}
{c^2[\rho^2+z^2+a^2-2a\rho\cos{(\phi-wt_{\rm r}})]},
\end{align}
where $t_{\rm r}=t-\sqrt{\rho^2+z^2+a^2-2a\rho\cos{(\phi-wt_{\rm r}})}/c$ is the retarded time. Using (E4) the first integral required in (E3) vanishes $\int_{V}\rho{\cal B}^v_\rho\, d\rho d\phi dz=0$. The second integral required in (E3) takes the form
\begin{equation}
\int_{V}\rho{\cal B}^a_\rho\, d\rho d\phi dz=\frac{qw^2a}{c^2}\int^{+\infty}_{-\infty}z dz\int_{R}^{\infty}\rho d\rho\oint_C\frac{\sin{(\phi-wt_{\rm r}})}{\rho^2+z^2+a^2-2a\rho\cos{(\phi-wt_{\rm r}})}d\phi.
\end{equation}
The third integral required in (E3) reads
\begin{equation}
\int_{V}z{\cal B}^v_z\, d\rho d\phi dz=-qwa\int^{+\infty}_{-\infty}zdz\int_{R}^{\infty}\rho d\rho\oint_C\frac{\rho\cos{(\phi-wt_{\rm r})-a}}{\rho^2+z^2+a^2-2a\rho\cos{(\phi-wt_{\rm r}})}d\phi.
\end{equation}
The four integral required in (E3) contains two pieces
\begin{align}
\int_{V}z{\cal B}^a_z\, d\rho d\phi dz=&\;\frac{qw^3a^2}{c^2}\int_{R}^{\infty}\rho d\rho\oint_C d\phi\int^{+\infty}_{-\infty}\frac{z[\rho\cos{(\phi-wt_{\rm r}})-a]^2}
{[\rho^2+z^2+a^2-2a\rho\cos{(\phi-wt_{\rm r}})]^{3/2}}dz \nonumber\\
&\;-\frac{qw^2a}{c^2}\int^{+\infty}_{-\infty}z dz\int_{R}^{\infty}\rho d\rho\oint_C\frac{\sin{(\phi-wt_{\rm r}})}{\rho^2+z^2+a^2-2a\rho\cos{(\phi-wt_{\rm r}})}d\phi.
\end{align}
The azimuthal integrals in (E8)-(E10) may be calculated at a fixed retarded time via a contour integration. Here we use the Mathematica program to calculate them
\begin{align}
\int_0^{2\pi n}\frac{\sin{(\phi-wt_{\rm r}})}{\rho^2+z^2+a^2-2a\rho\cos{(\phi-wt_{\rm r}})}d\phi =&\;0,\\
\int_0^{2\pi n}\frac{\rho\cos{(\phi-wt_{\rm r})-a}}{\rho^2+z^2+a^2-2a\rho\cos{(\phi-wt_{\rm r}})}d\phi =&\;0,
\end{align}
where $n$ is the winding number of the charge path. At a fixed retarded time, the first z-integral in (E10) is
\begin{equation}
\int^{+\infty}_{-\infty}\frac{z[\rho\cos{(\phi\!-\!wt_{\rm r}})-a]^2}
{[\rho^2\!+\!z^2\!+\!a^2\!-\!2a\rho\cos{(\phi\!-\!wt_{\rm r}})]^{3/2}}dz\!=\!0.
\end{equation}
Therefore we have $\int_{V}\rho{\cal B}^v_\rho\, d\rho d\phi dz\!=\!0, \int_{V}\rho{\cal B}^a_\rho\, d\rho d\phi dz\!=\!0,$
$ \int_{V}z{\cal B}^v_\rho\, d\rho d\phi dz=0$ and $\int_{V}z{\cal B}^a_\rho\, d\rho d\phi dz=0$
which imply the vanishing of (E3), and this means that $\bfL_{\rm R}=0$ for the charge-solenoid configuration.

\renewcommand\theequation{F\arabic{equation}}
\setcounter{equation}{0}
\section*{Appendix F Proof of (26)}
Consider the following representation of the Dirac delta function in azimuthal coordinates:
\begin{equation}
\delta\{ \phi' - \phi_{q}(t)\}= \frac{1}{2 \pi} \sum_{m=-\infty}^{\infty} {\rm e}^{im[\phi'- \phi_{q}(t)]},
\end{equation}
where $m$ is a real integer. Let us integrate this expression in a closed loop
\begin{equation}
\oint_{C}\delta\{ \phi' - \phi_{q}(t)\}\,d\phi'= \frac{1}{2 \pi} \sum_{m=-\infty}^{\infty} \oint_{C}{\rm e}^{im[\phi' - \phi_{q}(t)]}\,d\phi'.
\end{equation}
where $C$ denotes the charge path. If $C$ winds $n$ times around the solenoid then $\oint_{C} d\phi'= \int_{0}^{2\pi n}d\phi'$ where $n$ is the associated winding number and therefore (F2) reads
\begin{equation}
\int^{2\pi n}_{0}\delta\{ \phi' - \phi_{q}(t)\}\,d\phi'= \frac{1}{2 \pi}\sum_{m=-\infty}^{\infty} \int^{2\pi n}_{0}{\rm e}^{im[\phi'-  \phi_{q}(t)]}\,d\phi'.
\end{equation}
The integral in the right-hand side of (F3) gives
\begin{equation}
\int^{2\pi n}_{0}{\rm e}^{im[\phi' - \phi_{q}(t)]}\,d\phi'= \frac{2\sin(\pi mn)}{m} {\rm e}^{im[ n \pi - \phi_{q}(t)]},
\end{equation}
while the sum of this quantity over $m$ yields
\begin{equation}
\sum_{m=-\infty}^{\infty}\frac{2\sin(\pi mn)}{m} {\rm e}^{im[ n \pi - \phi_{q}(t)]}=2\pi n.
\end{equation}
Using (F3)-(F5) we obtain (26).

\end{document}